\documentclass[cits]{PoS}

\title{The accretion regime of LS~5039: 3-D SPH simulations}

\ShortTitle{Wind accretion in LS~5039}

\author{\speaker{A.T. Okazaki},$^a$ G.E. Romero,$^b$ and S.P. Owocki$^c$\\
        \llap{$^a$} Faculty of Engineering, Hokkai-Gakuen University, 
         Toyohira-ku, Sapporo 062-8605, Japan \\
        \llap{$^b$} Facultad de Ciencias Astron\'omicas y Geof\'{\i}sicas, 
         Universidad Nacional de La Plata, Paseo del Bosque, 1900 La Plata, 
         Argentina \\
        \llap{$^c$} Bartol Research Institute, University of Delaware, 
         Newark, DE 19716, USA \\
        E-mail: \email{okazaki@elsa.hokkai-s-u.ac.jp},
                \email{romero@fcaglp.unlp.edu.ar},
                \email{owocki@bartol.udel.edu}}

\abstract{LS 5039 is a TeV gamma-ray binary with extended radio emission. 
It consists of a compact object in the mildly eccentric (e=0.35), 3.9-day orbit 
around a massive O star. 
The nature of the compact object is not yet established.
In this paper, assuming that the compact object is a black hole,
we study the accretion of O-star wind by the black hole, 
by performing three-dimensional Smoothed Particle Hydrodynamics (SPH) simulations. 
In order to roughly emulate the effect of the stellar radiation effectively
canceling the stellar gravity, we assume that the O star's gravity does not
exert on the wind. The wind particles are ejected with
half the observed terminal velocity 
in a narrow range of azimuthal and vertical angles toward the black hole,
in order to emulate the wind significantly slower than the terminal speed,
and optimize the resolution and computational efficiency of simulations.
We find that the mass-accretion rate closely follows the classical Bondi-Hoyle-Littleton 
accretion rate, which is of the order of $10^{16}\,\mathrm{g\,s}^{-1}$ 
around periastron. 
The accretion rate at this level would provide jets enough power to produce the gamma-rays detected by HESS. Since the accretion peak occurs near the periastron passage, we need a strong gamma-ray absorption around periastron in order for the microquasar scenario to be consistent with the observed orbital modulation of the TeV gamma-ray flux.
}

\FullConference{7th INTEGRAL Workshop\\
		 September 8-11 2008\\
		 Copenhagen, Denmark}

\begin{document}

\section{Introduction}

The high mass X-ray binary LS~5039 consists of an O6.5V star and a compact object
in the $\sim 3.9$-day orbit with a moderate eccentricity ($e \sim 0.35$). 
It is one of the three binaries from which persistent TeV $\gamma$-rays
have been detected \cite{aha06}. 
The VHE $\gamma$-ray luminosity from LS~5039 modulates between $4-10 \times 10^{33}\,\mathrm{erg\,s}^{-1}$
\cite{aha06}. The maximum flux takes place around phase 0.8-0.9
\cite{aha06, bos05, cas05},
where phase 0 corresponds to the periastron passage.
The system shows a persistent radio emission, the morphology of which 
is consistent with mildly relativistic jets \cite{par00, par02}.
The nature of the compact object has not been established.
\cite{cas05} proposed that it is a black hole, which would place LS~5039
in the microquasar class, whereas some argue that the system is a colliding-wind binary 
with a non-accreting pulsar (e.g., \cite{dub06, mar05, tor07}).
To test the microquasar scenario for LS~5039, it is important to study 
whether the black hole can get high enough mass-accreion rate to power jets 
to produce the $\gamma$-rays detected by HESS.
In this paper, we briefly report on the result from 
three dimensional, numerical simulations of wind accretion by a black hole in LS~5039.

\section{Numerical model}

Simulations are performed with a 3-D Smoothed
Particle Hydrodynamics (SPH) code. The code is based on a version 
originally developed by Benz \cite{ben90a, ben90b}, 
then by Bate and his collaborators \cite{bat95}, and recently by 
\cite{oka08} and \cite{rom07}. 
It uses the variable smoothing length and the SPH equations with
the standard cubic-spline kernel are integrated with individual time step for
each particle. 
The artificial viscosity is assumed to have the standard values of
$\alpha_{\rm SPH}=1$ and $\beta_{\rm SPH}=2$. 
In our code, the O-star wind is modeled by
an ensemble of isothermal gas particles of negligible mass, 
while the compact object, which we assume to be a black hole, 
by a sink particle with the corresponding mass and a radius $r_\mathrm{acc}$
taken to be much smaller
than the classical Bondi-Hoyle-Littleton (BHL) accretion radius, $r_\mathrm{BHL}$, at periastron. 
If gas particles
fall within this radius, they are removed from the simulation.
On the other hand, we model the O star as a particle with the corresponding mass,
whose gravity exerts only on the binary companion, not on the wind particles. 
We take this assumption to roughly emulate the effect of 
the stellar radiation effectively canceling the stellar gravity.
Hence, in our simulation, the gravity by the compact object is the only external force
exerting on the wind particles.
The wind particles have the initial velocity of half the observed terminal velocity,
given that the accretion onto the black hole occurs in the region where
the wind speed is still significantly slower than the terminal velocity.
We set the binary orbit in the $x$-$y$ plane and the major axis of the orbit  
along the $x$-axis, where the apastron is in the $+x$-direction. 
The stellar, wind, and orbital parameters adopted for LS~5039 simulations are summarized 
in Table~\ref{tbl:params}.

\begin{table}[!ht]
\caption{Model parameters for LS~5039}
\centerline{
\begin{tabular}{lcc}
\hline
 & Primary & Secondary \\
\hline
\hline
Spectral type & O6.5V & Black Hole \\
Mass ($M_{\odot}$) & 22.9$^{\rm a}$ & 3.7$^{\rm a}$ \\
Radius & $9.3 R_{\odot}$$^{\rm a}$ ($=0.31a$) & $2.5 \times 10^{-3}a$ \\
Effective temperature $T_\mathrm{eff}$ (K) & 39,000$^{\rm a}$ & -- \\
Initial wind velocity $V_\mathrm{w,ini}$ (${\rm km}\,{\rm s}^{-1}$) &
   1,200 & -- \\
Mass loss rate $\dot{M}_{*}$ ($M_{\odot}\,{\rm yr}^{-1}$) & 
   $5 \times 10^{-7}$\,\,$^{\rm b}$ &  -- \\
Orbital period $P_{\rm orb}$ (d) & \multicolumn{2}{c}{3.9060$^{\rm a}$} \\
Orbital eccentricity $e$ & \multicolumn{2}{c}{0.35$^{\rm a}$} \\
Semi-major axis $a$ (cm) & \multicolumn{2}{c}{$2.17 \times 10^{12}$}\\
\hline
\multicolumn{3}{l}{$^{\rm a}$ \cite{cas05}} \\
\multicolumn{3}{l}{$^{\rm b}$ The upper limit of the low X-ray state and
   the lower limit of the high state \cite{cas05}.}
\end{tabular}}
\label{tbl:params}
\end{table}

\begin{figure}[!ht]
\centerline{
\includegraphics*[width=0.35\textwidth]{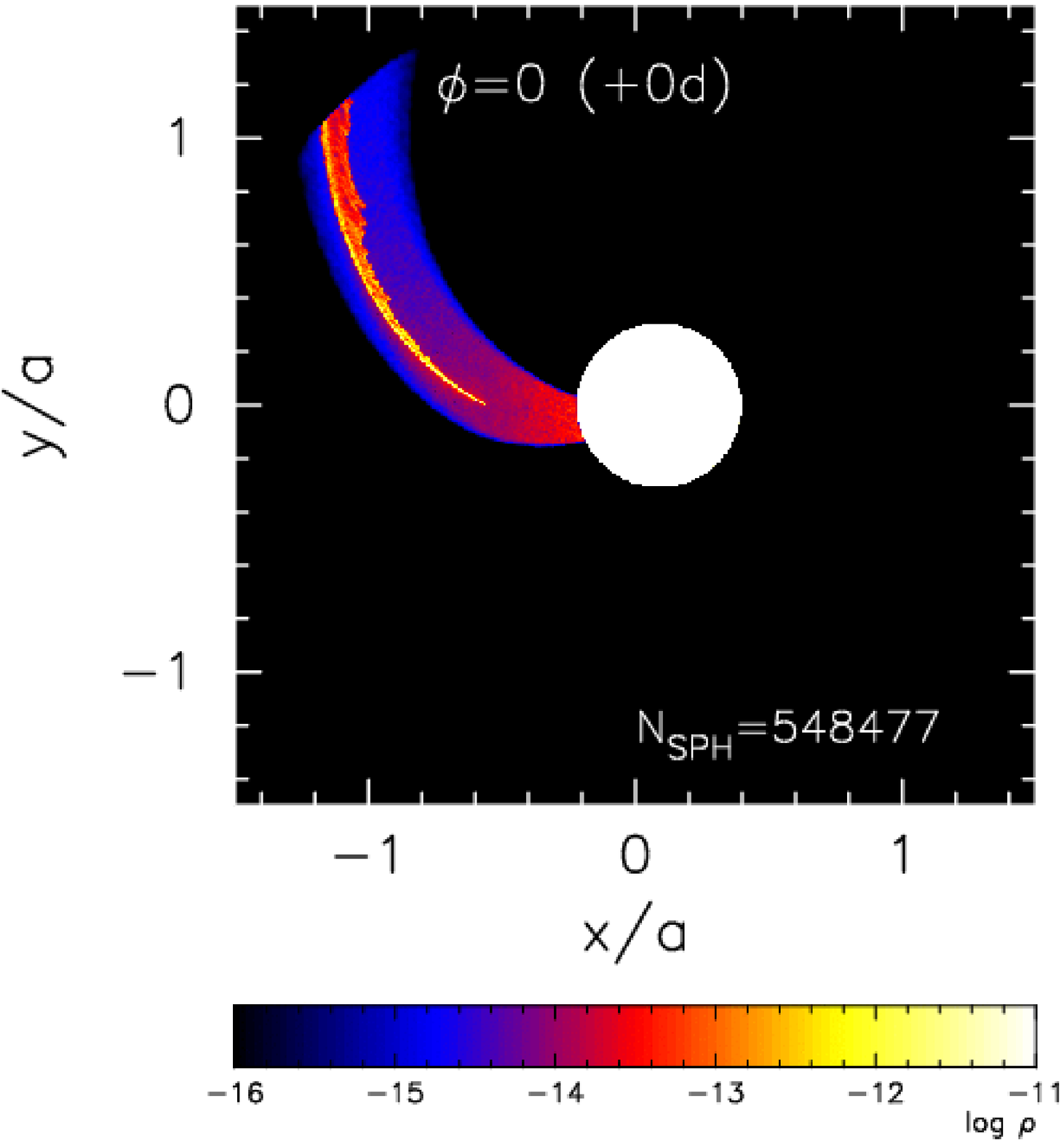}
\includegraphics*[width=0.35\textwidth]{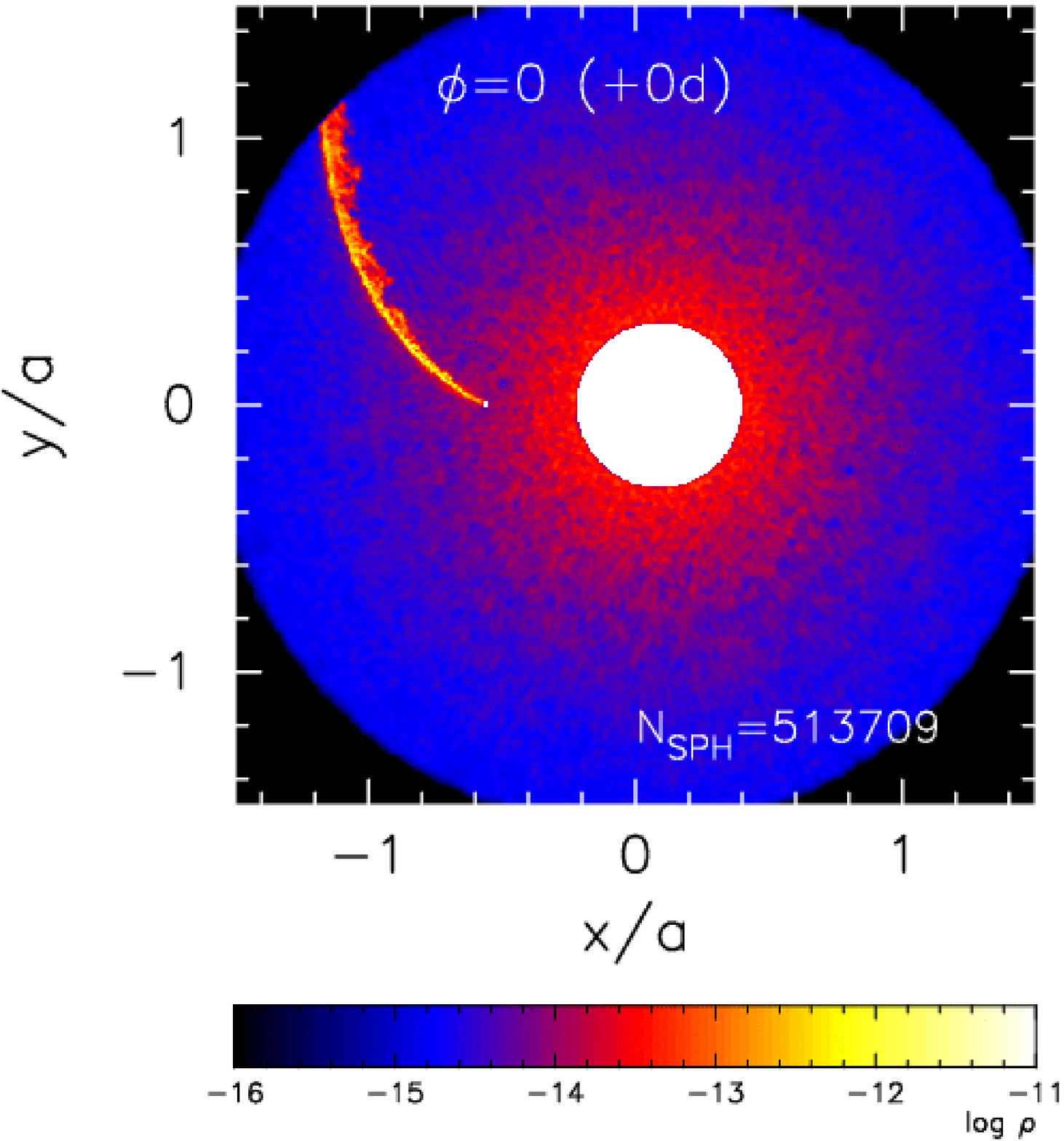}
}
\caption{Wind density in the orbital plane at periastron. 
The left panel shows the logarithmic density from a simulation where wind particles 
are ejected  toward the black hole (a small bright dot
on the left of the O star) in the restricted range of $30^{\circ}$ 
in the azimuthal direction and $20^{\circ}$ in the polar direction, 
whereas the right panel shows that from a simulation of an axisymmetric wind with the same
opening angle of $20^{\circ}$. Annotated in each panel are the phase, the time 
from the periastron passage, and the number of wind particles.
$r_\mathrm{acc}=2.5 \times 10^{-3}$ in the left panel, while 
$r_\mathrm{acc}=10^{-2}$ in the right panel because of the lower spatial resolution.}
\label{fig:comp}
\end{figure}

To optimize the resolution and computational efficiency of our simulations, 
the wind particles are ejected only in a narrow range of azimuthal and vertical angles 
toward the black hole
\footnote{In the poster presented at the Workshop, we had broken the stellar wind and accretion portions of the computation into two separate, but linked parts, in which 
the first focused on the stellar wind expansion and the rate of mass that 
enters a spherical region around the balck hole, and the second focused
the mass accretion onto the black hole, using the information of 
the first part of the simulation.
However, after the Workshop, it turned out that the second part had a bug that seriously affected 
the final result. Therefore, we have revised our code and this time rerun simulations
of the whole system, using a much smaller accretion radius. The result reported 
in this paper is based on these revised simulations.}.
We have confirmed that this method provides a quite similar flow structure to
that of a corresponding spherically symmetric wind. Moreover, the former 
has a much higher spatial resolution than the latter, as expected. We can see this in 
Fig.~\ref{fig:comp} comparing the density distribution in the orbital plane at periastron 
from such a simulation with that from a simulation of an axisymmetric wind ejected within 
a constant opening angle of $20^{\circ}$. Both simulations use similar number of 
wind particles $N_\mathrm{SPH}$. 
In the former simulation, the accretion radius of the black hole 
$r_\mathrm{acc}=2.5 \times 10^{-3}$, while 
$r_\mathrm{acc}=10^{-2}$ for the latter simulation because of the lower spatial resolution.

\section{Results}

\begin{figure}[!ht]
\centerline{
\begin{tabular}{c@{}c@{}c}
\includegraphics*[width=0.33\textwidth]{CSWIND150001.eps} &
\includegraphics*[width=0.33\textwidth]{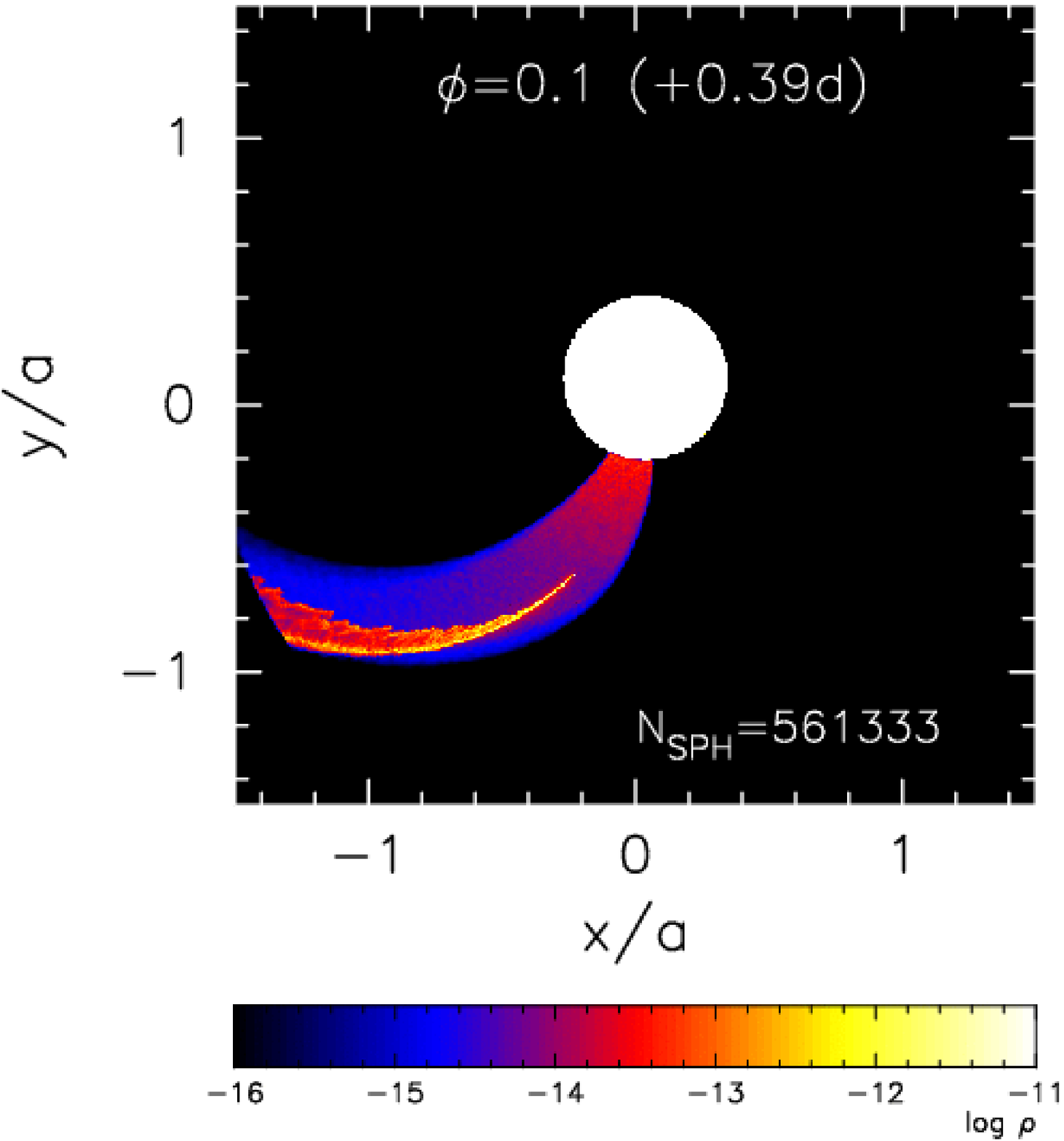} &
\includegraphics*[width=0.33\textwidth]{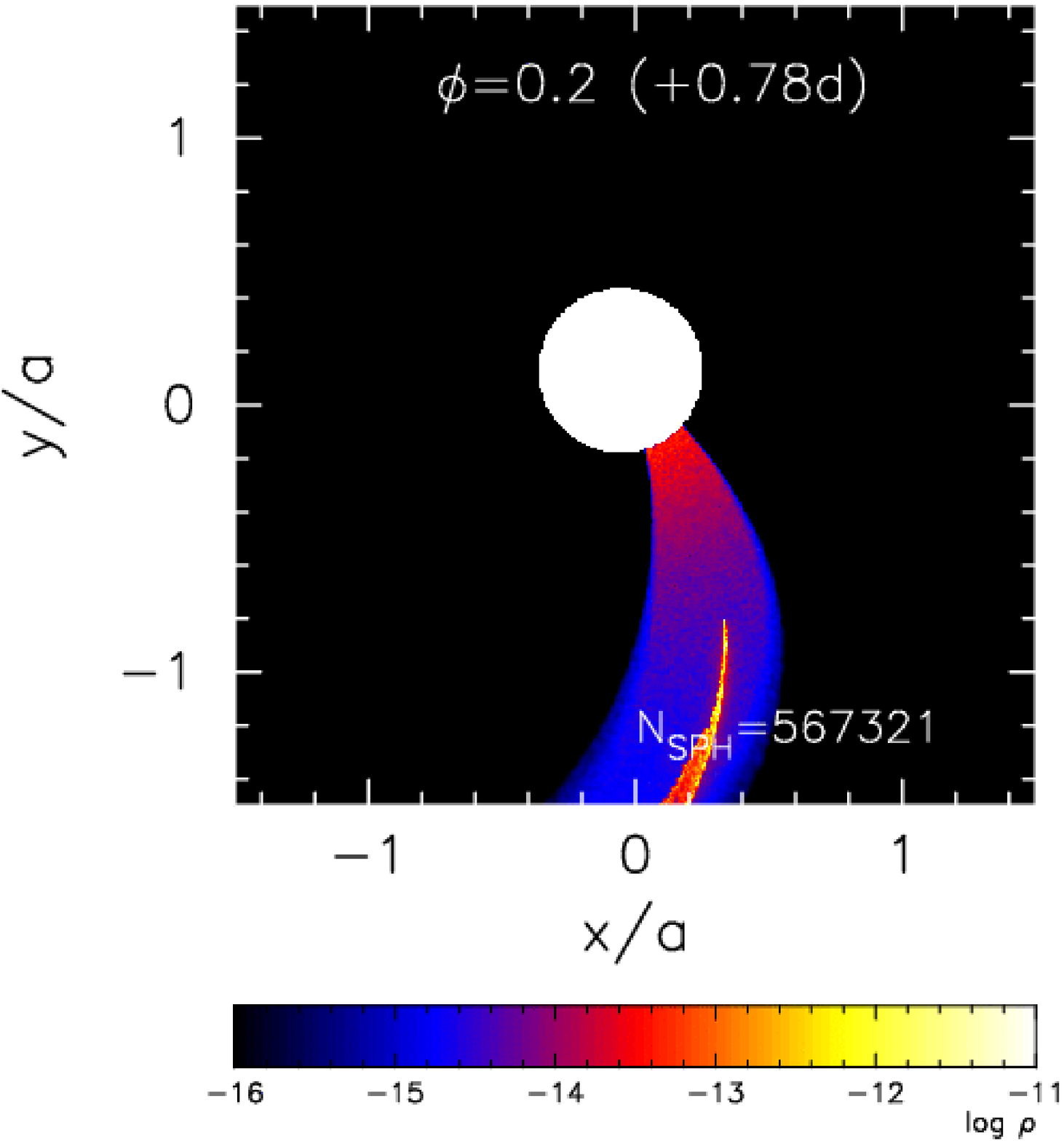} \\
\includegraphics*[width=0.33\textwidth]{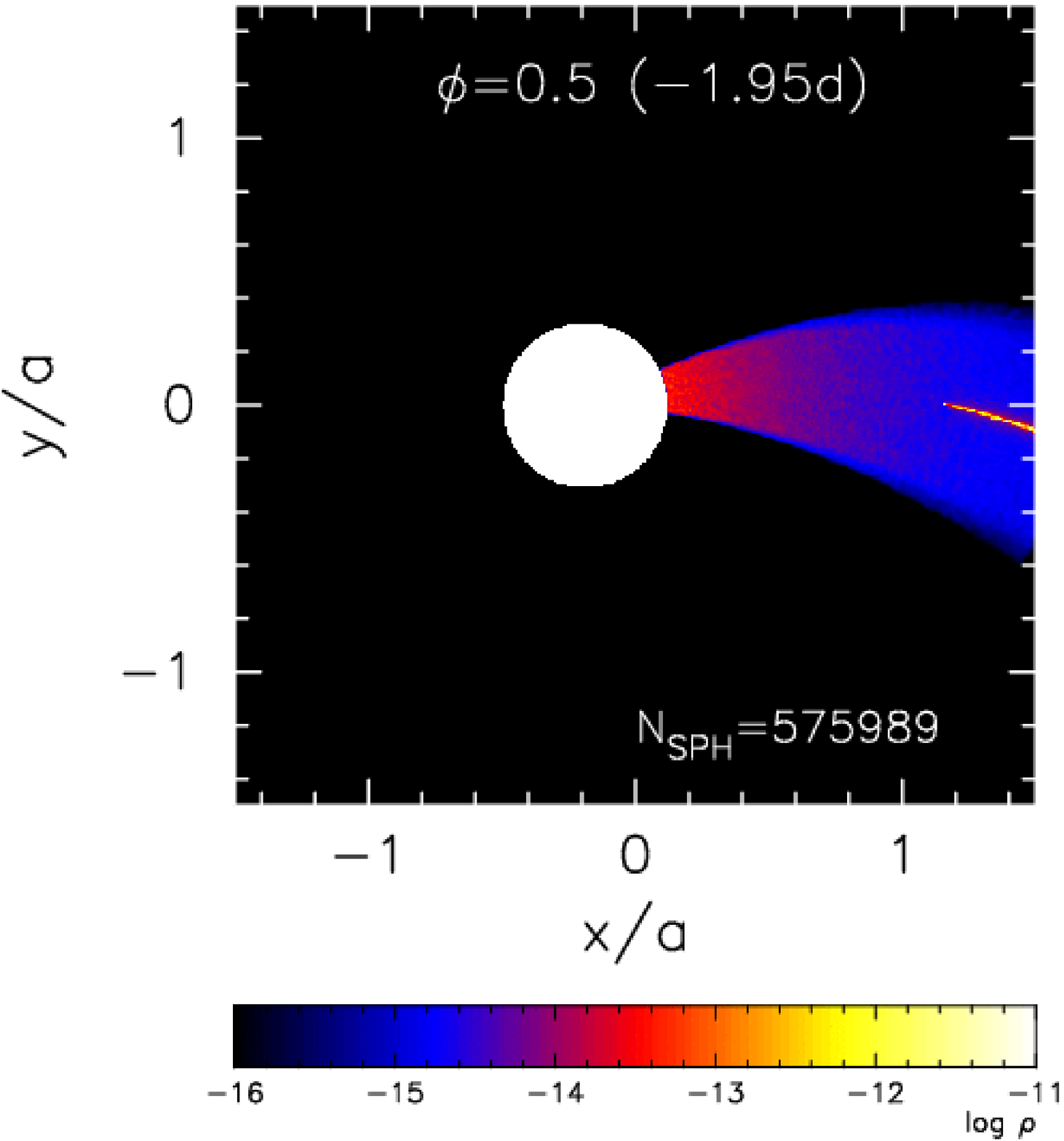} &
\includegraphics*[width=0.33\textwidth]{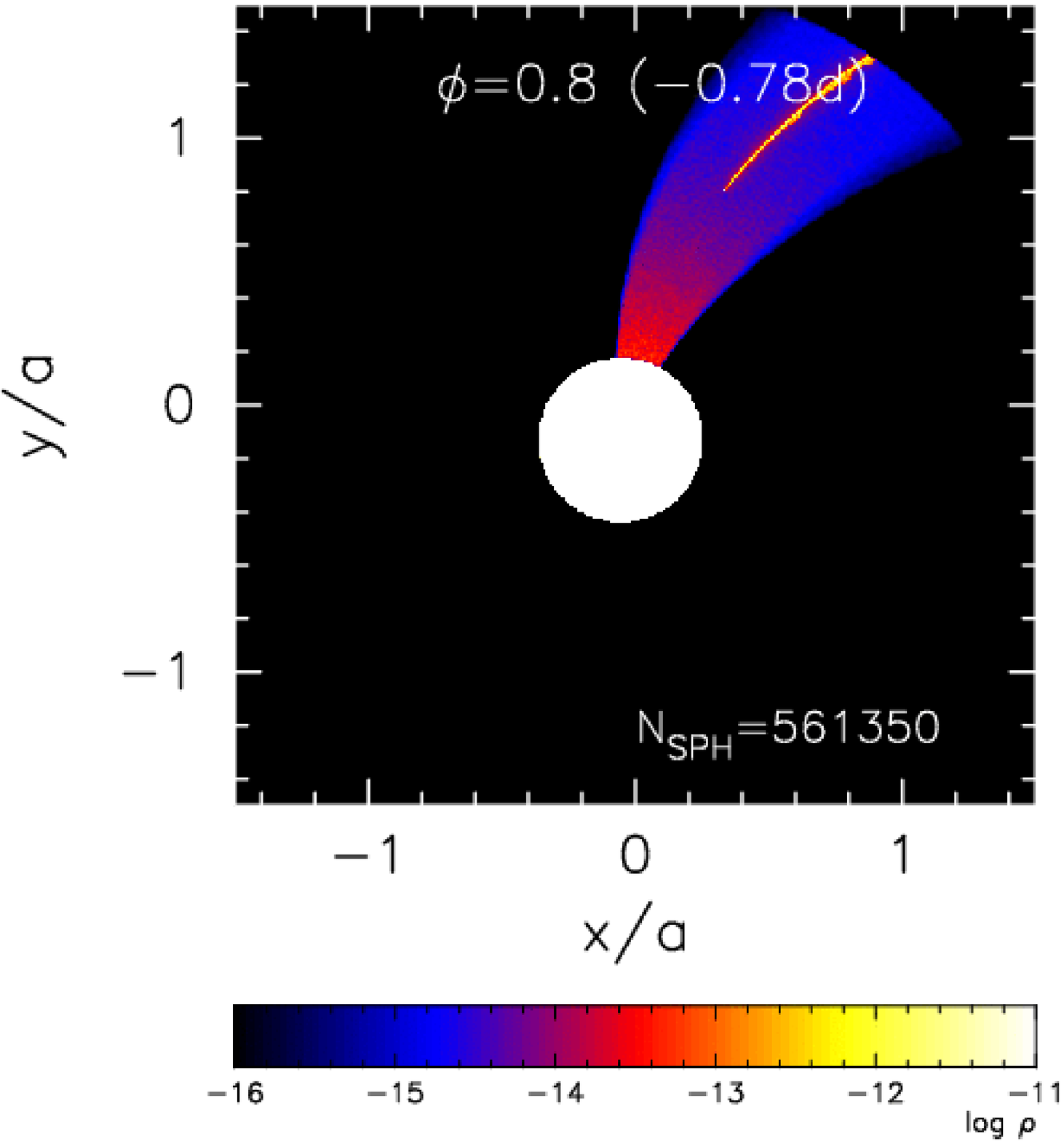} &
\includegraphics*[width=0.33\textwidth]{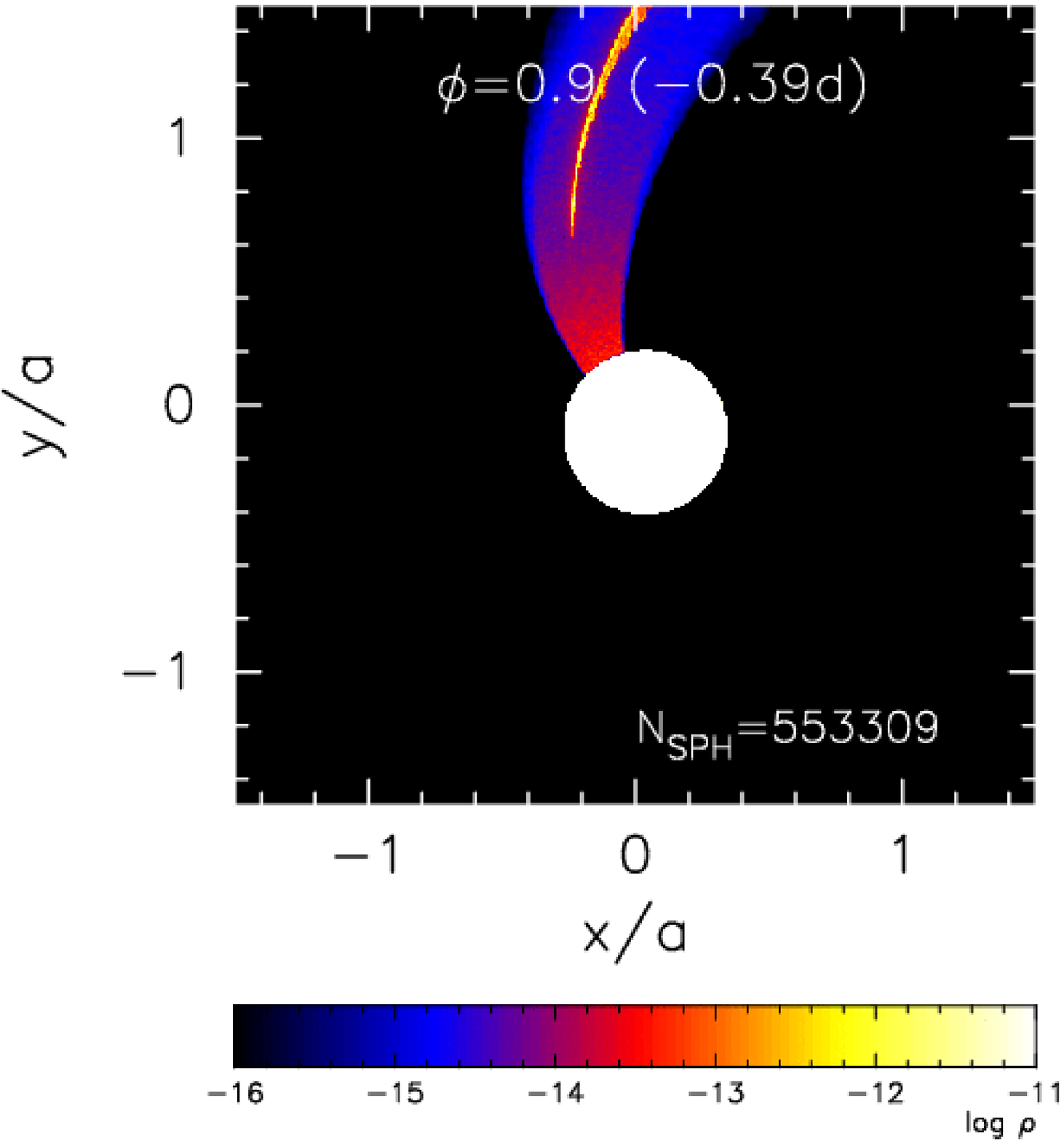}
\end{tabular}
}
\caption{Snapshots at six different phases.
The color-scale plot shows the logarithmic density in the orbital plane in cgs units.
The large filled circle near the center is the O star, 
while a tiny bright dot (with the radius of $2.5 \times 10^{-3}a$) 
at the tip of bow shocks is the black hole.}
\label{fig:snapshots1}
\end{figure}

\begin{figure}[!ht]
\centerline{
\begin{tabular}{c@{}c@{}c}
\includegraphics*[width=0.33\textwidth]{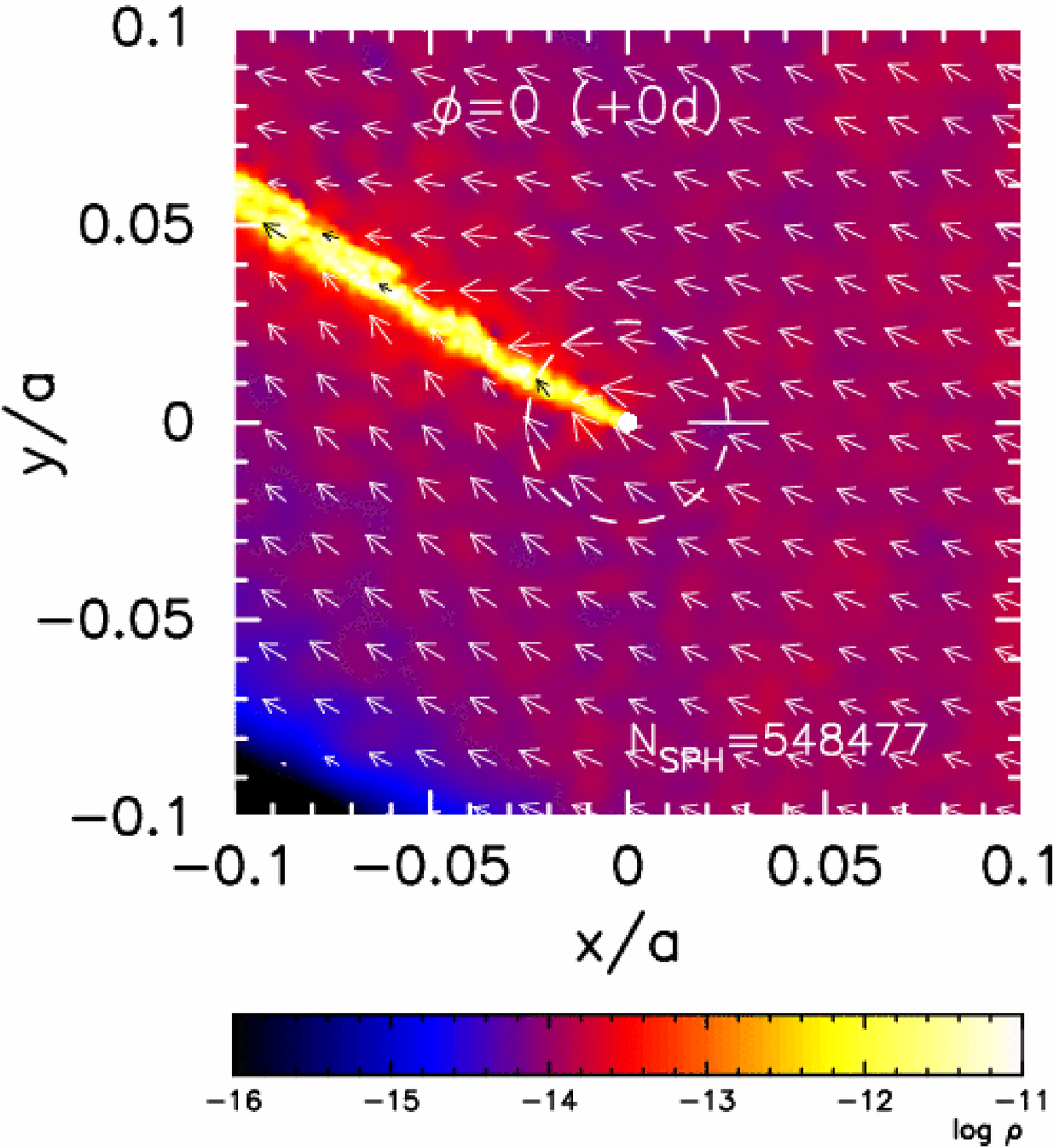} &
\includegraphics*[width=0.33\textwidth]{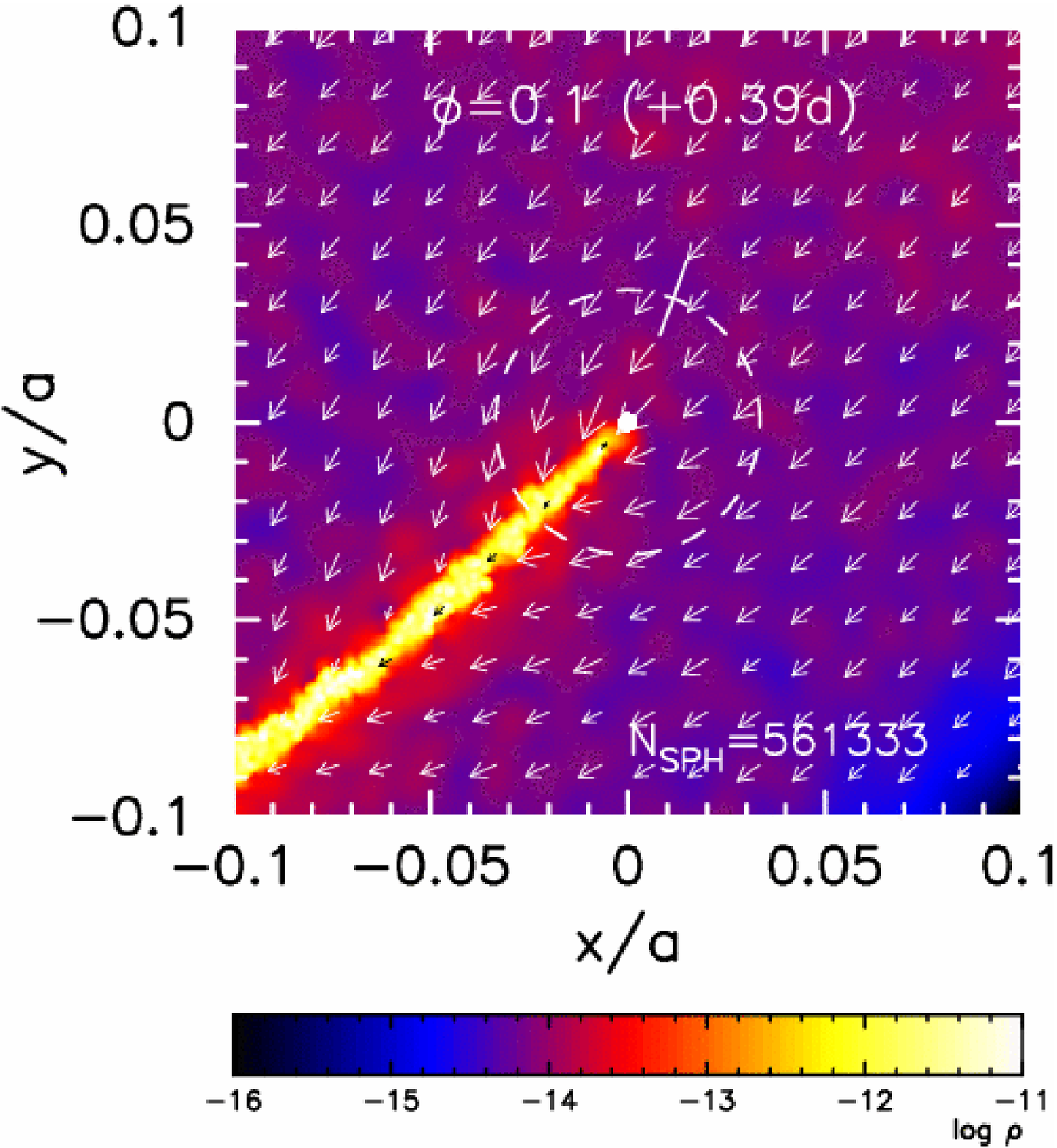} &
\includegraphics*[width=0.33\textwidth]{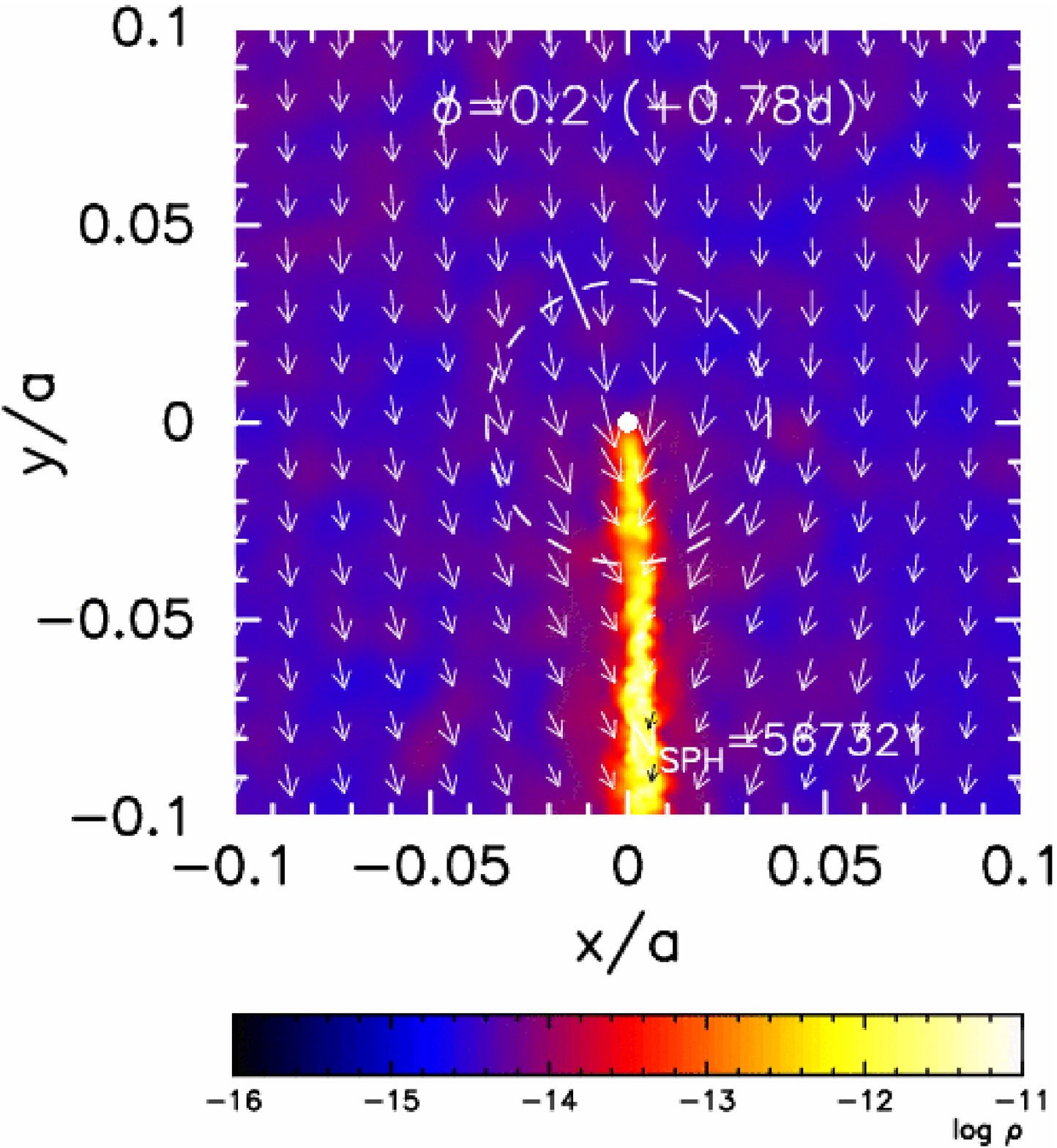} \\
\includegraphics*[width=0.33\textwidth]{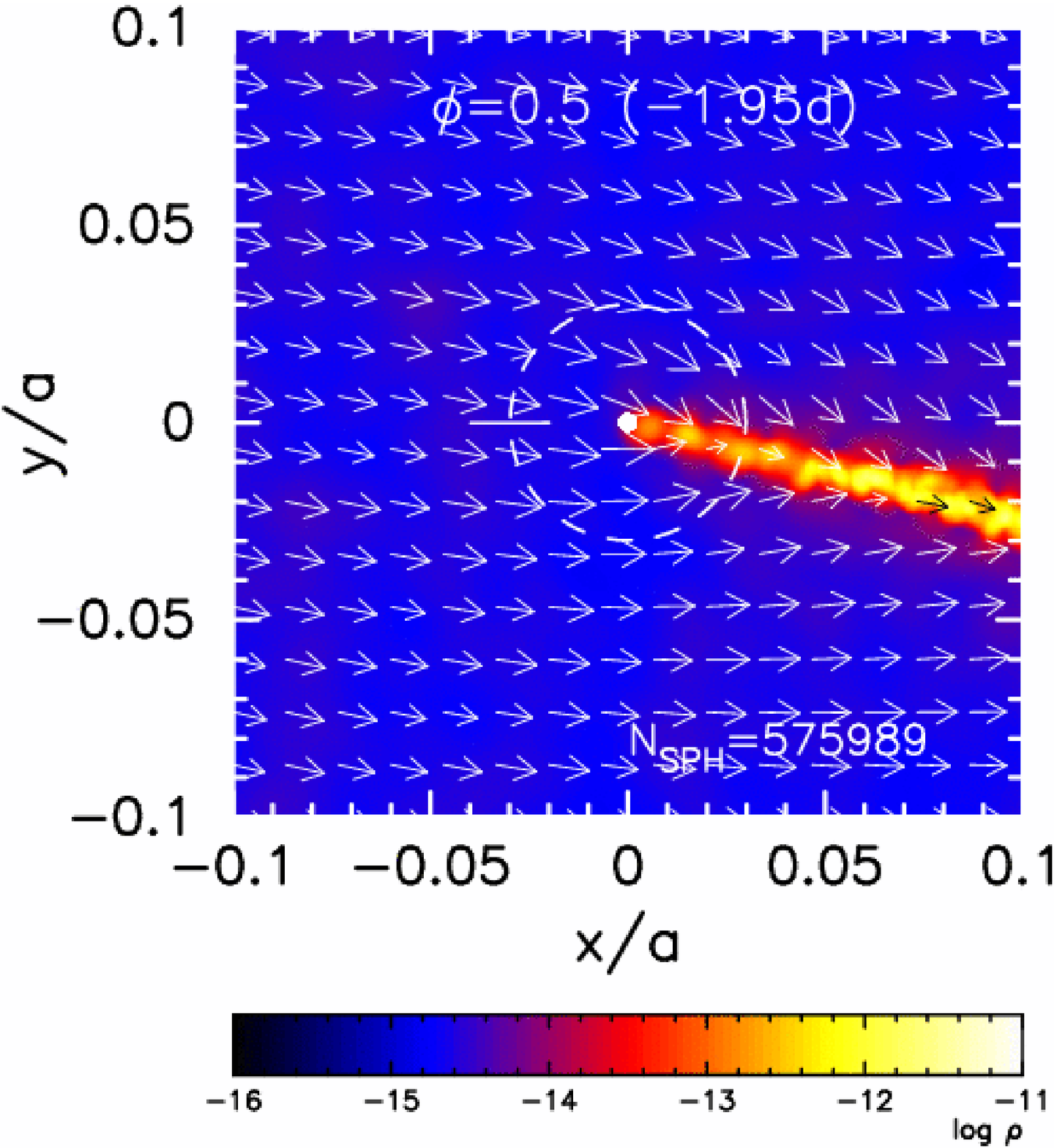} &
\includegraphics*[width=0.33\textwidth]{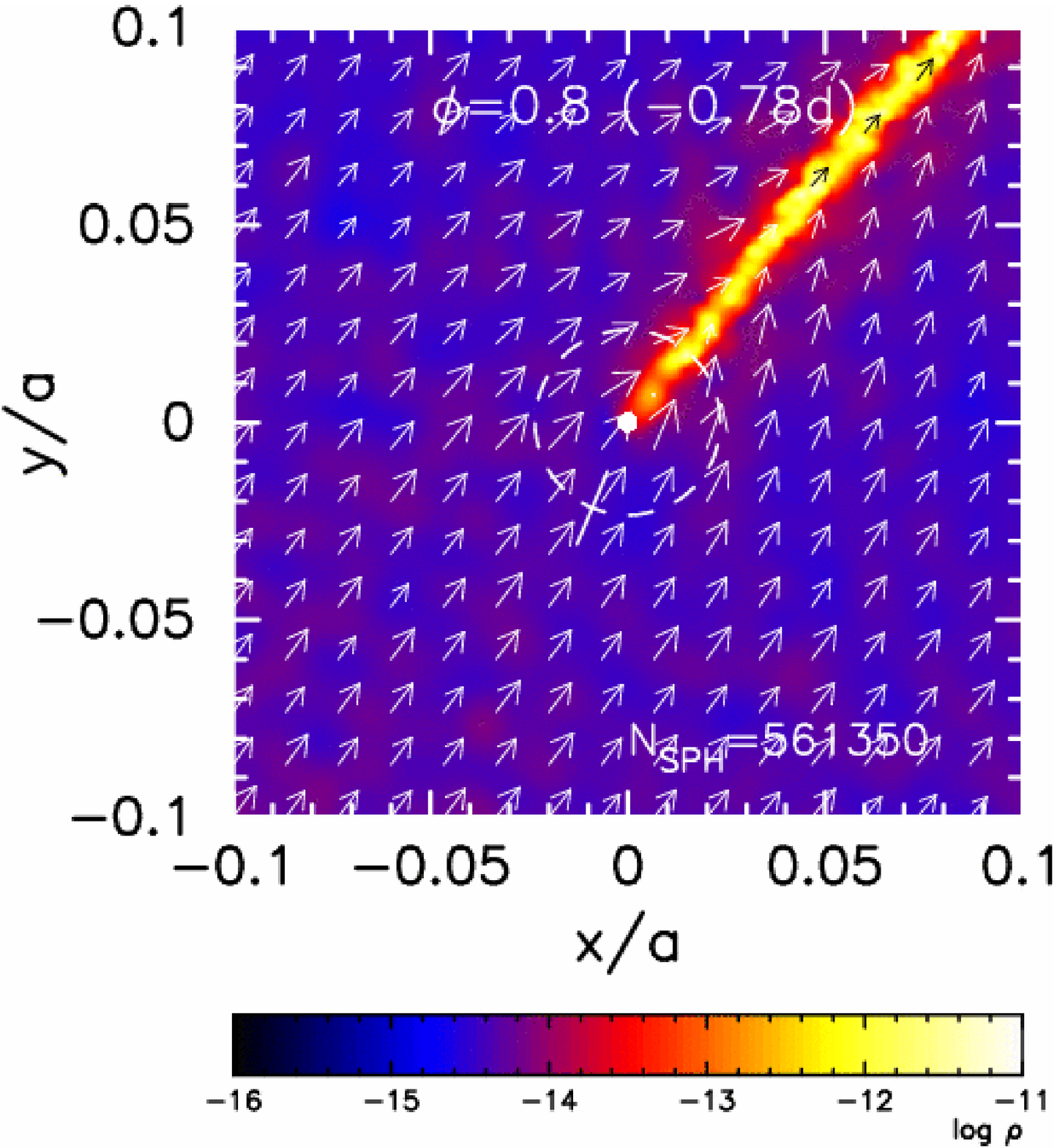} &
\includegraphics*[width=0.33\textwidth]{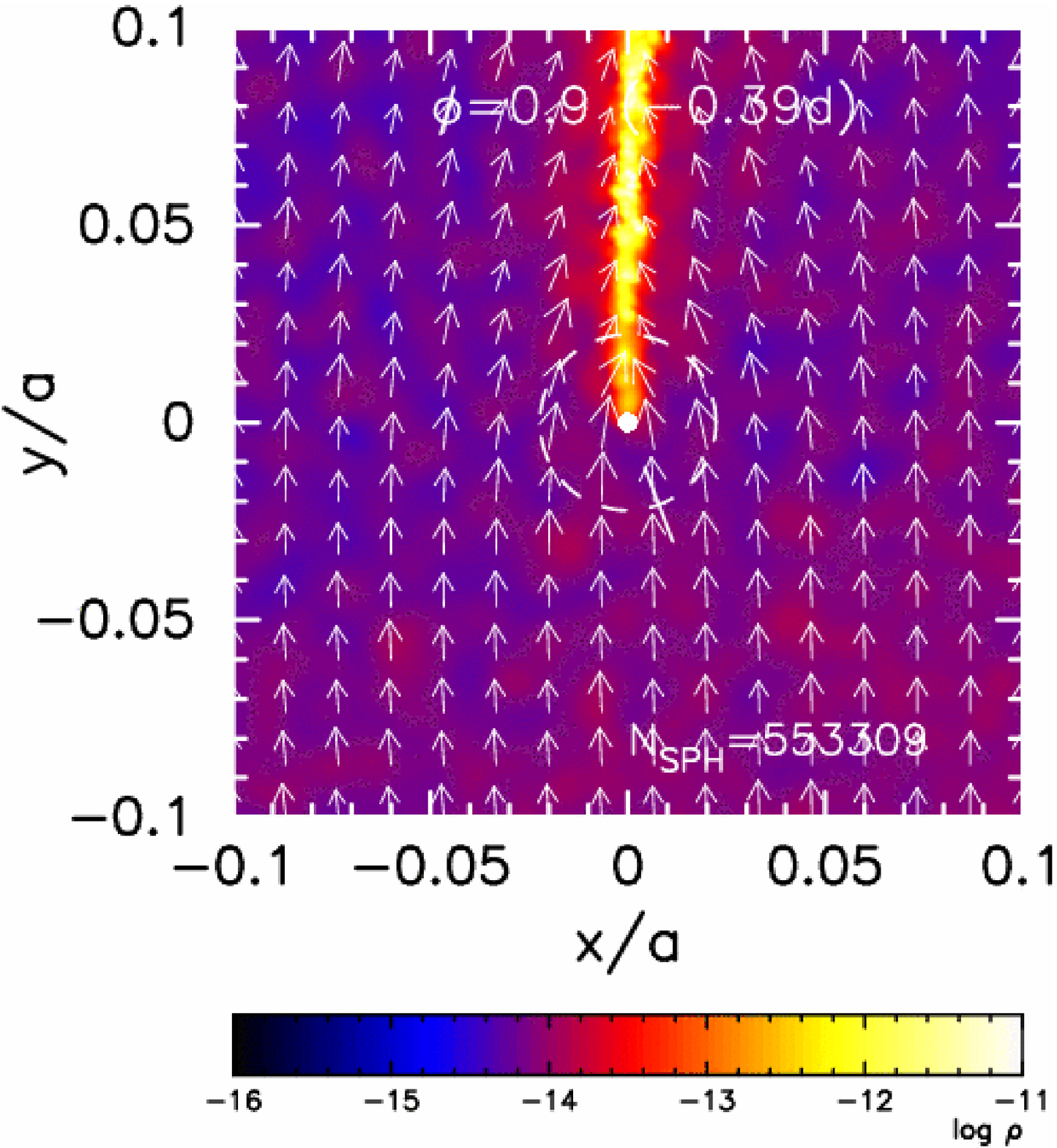}
\end{tabular}
}
\caption{Detailed flow structure around the black hole at the same phases 
as in Fig.~2.
Color-scale plot shows the logarithmic density in the orbital plane, 
while arrows denote the velocity vectors of the flow.
The dashed circle around the black hole denotes 
the Bondi-Hoyle-Littleton accretion radius, $r_\mathrm{BHL}$, 
for a $1,200\,\mathrm{km\,s}^{-1}$ wind.
The line fragment on the dashed circle shows the direction of the O star.
}
\label{fig:snapshots2}
\end{figure}

In Figs.~\ref{fig:snapshots1} and \ref{fig:snapshots2}, 
we present the density distribution of the wind in 
the binary orbital plane at several different phases
($\phi=0$ corresponds to periastron).
Figure~\ref{fig:snapshots1} covers the whole simulation volume
(within $1.5a$ from the center of mass of the binary), while 
Fig.~\ref{fig:snapshots2} shows the detailed structure of the wind 
around the black hole.
In both figures, the color-scale plots show the logarithmic density of the wind
in the orbital plane, while arrows superposed on the density plot 
in Fig.~\ref{fig:snapshots2} denote the velocity vectors of the flow.
In Fig.~\ref{fig:snapshots2}, the dashed circle denotes 
the classical BHL accretion radius, 
$r_\mathrm{BHL}$, for a supersonic flow, which is given by
\begin{equation}
   r_\mathrm{BHL} = \frac{2GM_\mathrm{X}}{v^2_\mathrm{rel}},
\label{eq:r_bhl}
\end{equation}
where $M_\mathrm{X}$ is the mass of the black hole and $v_\mathrm{rel}$ is
the relative velocity between the wind and the black hole (e.g., \cite{fra02}).
Here we have calculated $r_\mathrm{BHL}$ by taking 
$v_\mathrm{w, ini}=1,200\,\mathrm{km\,s}^{-1}$ as the wind velocity.
The line fragment on the dashed circle in Fig.~\ref{fig:snapshots2} shows the direction of the O star.

Figure~\ref{fig:snapshots1} shows that strong bow shocks form behind the black hole 
and that its shape is distorted around periastron by the rapid orbital motion
of the black hole.
Note that, despite the spiral shape of the wind region, 
which is a direct consequence of our wind ejection method, 
the flow is radial except in a region near bow shocks.

\begin{figure}[!ht]
\centerline{
\includegraphics*[width=0.4\textwidth]{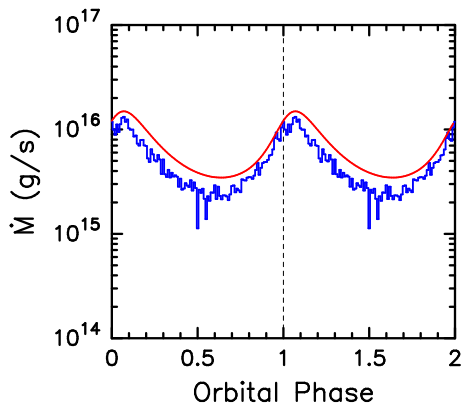}
}
\caption{Mass-accretion rate onto the black hole, which is averaged over
   the phase interval of 0.01. For comparison, the BHL accretion rate for
   a wind of $1,200\,\mathrm{km\,s}^{-1}$ is shown by the red line.}
\label{fig:mdot}
\end{figure}

From Fig.~\ref{fig:snapshots2}, we can see that 
the accretion is basically of classical BHL type.
The wind passing through a region within $\sim r_\mathrm{BHL}$ from the black hole is 
strongly focussed and ultimately accreted, while the flow is almost parallel in an outer region.
In order to see this more clearly, we compare in Fig.~\ref{fig:mdot} 
the simulated mass-accretion rate, $\dot{M}$, with 
the BHL accretion rate, $\dot{M}_\mathrm{BHL}$, given by
\begin{equation}
   \dot{M}_\mathrm{BHL} = \frac{G^2M^2_\mathrm{X}\dot{M}_{*}}
                               {v^4_\mathrm{rel}d^2},
\label{eq:mdot_bhl}
\end{equation}
where $d$ is the binary separation (e.g., \cite{fra02}).
Here we have calculated $v_\mathrm{rel}$, assuming the constant wind velocity of
$v_\mathrm{w,ini}$.
In this figure, we have averaged the rapidly fluctuating accretion rate 
over the phase interval of 0.01 (histogram in blue).
The red line denotes the BHL accretion rate, $\dot{M}_\mathrm{BHL}$.
It is immediately observed that the simulated accretion rate closely follows
the BHL rate. Thus, the classical BHL approximation also basically hold
in 3-D simulations as it does in 2-D counterparts.
A closer look at the figure, however, tells us that the simulated accretion rate is slightly lower 
than the BHL rate. This is due to the effect of wind acceleration by the gravity of black hole;
the assumption of constant wind velocity results in an slightly underestimated relative velocity, 
which in turn leads to a slightly overestimated BHL accretion rate shown in Fig.~\ref{fig:mdot}.
The simulated accretion rate modulates by a factor of $\sim 5$. 
It varies from $\sim 10^{16}\mathrm{g\,s}^{-1}$
around periastron to $\sim 2 \times 10^{15}\mathrm{g\,s}^{-1}$ around apastron.

It is important to note that the mass-accretion rate of $\sim 10^{16}\mathrm{g\,s}^{-1}$
would provide jets enough power to produce the $\gamma$-rays detected by HESS.
With $\dot{M} \sim 10^{16}\mathrm{g\,s}^{-1}$,
the accretion power is $\sim 10^{37}\,\mathrm{erg\,s}^{-1}$.
Assuming that 10\% of this power goes to jets and $\sim 1/3$ of the jet power is converted
to relativistic particles, then we expect a $\gamma$-ray luminosity of
roughly $\sim 3 \times 10^{35}\,\mathrm{erg\,s}^{-1}$, which would explain the observed
$\gamma$-ray luminosity of $2.7 \times 10^{35}\,\mathrm{erg\,s}^{-1}$ above 100\,MeV
\cite{har99}.

Now, with the accretion rate as high as $\sim 10^{16}\mathrm{g\,s}^{-1}$,
why do we see no thermal feature at X-rays (e.g., \cite{bos08})?
One possibility is that the accretion flow in LS~5039 is radiatively inefficient.
For a black hole of $3.7 M_{\odot}$, the Eddington luminosity is 
$\sim 4.6 \times 10^{38}\,\mathrm{erg\,s}^{-1}$. 
With the energy conversion rate $\sim 0.1$, the critical mass accretion rate,
$\dot{M}_\mathrm{crit}$ is $\sim 5 \times 10^{18}\,\mathrm{g\,s}^{-1}$. 
Then, the accretion rate of $\sim 10^{16}\mathrm{g\,s}^{-1}$ corresponds to
$\sim 2 \times 10^{-3}\dot{M}_\mathrm{crit}$. This rate is well below 
the maximum accretion rate, $0.05-0.1 \dot{M}_\mathrm{crit}$, below which the
advection-dominated accretion solutions exist \cite{nar98}.
Thus, our model is consistent with ADAF solutions, which could explain 
the paucity of thermal X-ray emission in LS~5039.

\section{Conclusions}
We have carried out 3-D SPH simulations of the wind accretion in LS~5039,
assuming that the compact object is a $3.7 M_{\odot}$ black hole.
Wind particles were ejected at 
half the observed terminal velocity to
emulate the stellar wind significantly lower than the terminal velocity.
In order to optimize the resolution and computational efficiency of simulations,
we have simulated a wind restricted to a narrow range of angles
toward the black hole.
We have found that the simulated accretion rate closely follows,
but is slightly lower than,
the classical Bondi-Hoyle-Littleton accretion rate.
The maximum accretion rate of $\sim 10^{16}\,\mathrm{g\,s}^{-1}$ 
for the stellar mass-loss rate of $5 \times 10^{-7} M_{\odot}\,\mathrm{yr}^{-1}$
takes place at a phase slightly after the periastron passage.
This rate would provide jets in LS~5039 enough power to produce 
the $\gamma$-rays detected by EGRET and HESS.
Since the accretion peak occurs near periastron, we need a strong $\gamma$-ray absorption around periastron in order for the microquasar scenario to be consistent with the observed peak phase of 0.8-0.9 of the TeV $\gamma$-ray flux.


\begin{thebibliography}{99}
  \bibitem{aha06}
     Aharonian F., et al. (HESS Collaboration), 
     \emph{3.9 day orbital modulation in the TeV gamma-ray flux and
     spectrum from the X-ray binary LS~5039}, A\&A {\bf 460}, 743 (2006)
  \bibitem{bat95}
     Bate M.R., Bonnell I.A. \& Price N.M., 
     \emph{Modelling accretion in protobinary systems}, MNRAS {\bf 277}, 362 (1995)
  \bibitem{ben90a}
     Benz W., in Buchler J. R.,ed., 
     \emph{The Numerical Modelling of Nonlinear Stellar 
     Pulsations}. Kluwer, Dordrecht, p.269 (1990)
  \bibitem{ben90b}
     Benz W., Bowers R.L., Cameron A.G.W. \& Press W.H., 
     \emph{Dynamic mass exchange in doubly degenerate binaries. 
     I - 0.9 and 1.2 solar mass stars}, ApJ {\bf 348}, 647 (1990)
  \bibitem{bos05} 
	Bosch-Ramon V., Paredes J.M., Rib\'{o} M., Miller J.M., Reig P., Mart\'{i} J.,
	\emph{Orbital X-Ray Variability of the Microquasar LS~5039}, 
	ApJ {\bf 628}, 388 (2005)
  \bibitem{bos08}
     Bosch-Ramon V., Motch C., Rib\'{o} M, Lopes de Oliveira R., Janot-Pacheco E.,
     Negueruela I., Paredes J. M., Martocchia A., 
     \emph{Exploring the connection between the stellar wind and 
     the non-thermal emission in LS~5039}, A\&A {\bf 473}, 545 (2007)
  \bibitem{cas05}
     Casares J., Rib\'{o} M., Ribas I., Paredes J. M., Mart\'{i} J., Herrero, A.,
     \emph{A possible black hole in the $\gamma$-ray microquasar LS~5039}, 
     MNRAS {\bf 364}, 899 (2005)
  \bibitem{dub06}
     Dubus G., \emph{Gamma-ray binaries: pulsars in disguise?},
     A\&A {\bf 456}, 801 (2006)
  \bibitem{fra02}
     Frank J., King A.R., \& Raine D.J., \emph{Accretion power in Astrophysics, 
     3rd edn}. Cambridge Univ. Press, Cambridge, pp.74, 75 (2002)
  \bibitem{har99}
     Hartman R.C., Bertsch, D.L., Bloom, S.D., et al., 
     \emph{The Third EGRET Catalog of High-Energy Gamma-Ray Sources},
     ApJS {\bf 123}, 79 (1999)
  \bibitem{mar05}
     Martocchia A., Motch C., \& Negueruela I., 
     \emph{The low X-ray state of LS~5039/RX J1826.2-14502005}, A\&A, {\bf 430}, 245 (2005)
  \bibitem{nar98}
     Narayan R., Mahadevan R., \& Quataert E., in: M.A. Abramowickz, 
     G. Bj\"ornsson, \& J.E. Pringle (eds), \emph{Theory of Black Hole Accretion Disks}, 
     Cambridge University Press, Cambridge, p.148 (1998)
  \bibitem{oka08}
     Okazaki A.T., Owocki S.P., Russell C.M.P. \& Corcoran M.F., 
     \emph{Modelling the RXTE light curve of $\eta$~Carinae from a 3D SPH simulation of 
     its binary wind collision}, 
     MNRAS Letters {\bf 388}, L39 (2008)
  \bibitem{par00}
     Paredes, J.M., Mart\'{i}, J., Rib\'{o}, M., \& Massi, M.,
     \emph{Discovery of a High-Energy Gamma-Ray-Emitting Persistent Microquasar},
     Science {\bf 288}, 2340 (2000)
  \bibitem{par02}
     Paredes, J.M., Rib\'{o}, M., Ros, E., Mart\'{i}, J., \& Massi, M.,
     \emph{Confirmation of persistent radio jets in the microquasar LS~5039},
     A\&A {\bf 393}, L99 (2002)
  \bibitem{rom07}
     Romero G.E., Okazaki A.T., Orellana M., \& Owocki S.P.,
     \emph{Accretion vs. colliding wind models for the gamma-ray binary LS I +61 303: 
     an assessment}, A\&A {\bf 474}, 15 (2007)
  \bibitem{tor07}
     Sierpowska-Bartosik A. \& Torres D.F., 
     \emph{Pulsar Model of the High-Energy Phenomenology of LS~5039},
     ApJ Letters {\bf 671}, 145 (2007)
\end{thebibliography}
\end{document}